\documentclass[intlimits,twoside,a4paper]{article}
\usepackage{wrapfig}
\usepackage{graphicx}
\usepackage[T2A]{fontenc}
\usepackage{amssymb,amsmath}

\usepackage[cp1251]{inputenc}
\usepackage{cmpj}

\issue{2011}{14}{1}{13801}

\doinumber{10.5488/CMP.14.13801}

\makeatletter
\newcounter{robA}
\newcounter{robB}
\newcounter{robC}
\newcounter{robD}
\newcounter{robE}
\newcounter{robF}
\newcounter{robG}

\def\trojkat(#1,#2,#3){
    \setcounter{robA}{#3}
    \divide\c@robA2
    \put(#1,#2){\line(1,0){#3}}
    \put(#1,#2){\line(2,3){\value{robA}}}
    \setcounter{robB}{#1}
    \addtocounter{robB}{#3}
    \put(\value{robB},#2){\line(-2,3){\value{robA}}}
}
\def\trojkatABC(#1,#2,#3){
    \trojkat(#1,#2,#3)
    \setcounter{robA}{#1}
    \addtocounter{robA}{-8}
    \setcounter{robB}{#2}
    \addtocounter{robB}{-5}
    \put(\value{robA},\value{robB}){$a$}
    \setcounter{robC}{#3}
    \addtocounter{robC}{3}
    \addtocounter{robC}{#1}
    \put(\value{robC},\value{robB}){$b$}
    \addtocounter{robA}{10}
    \setcounter{robD}{#3}
    \divide\c@robD2
    \addtocounter{robA}{\value{robD}}
    \divide\c@robD2
    \multiply\c@robD3
    \addtocounter{robD}{2}
    \addtocounter{robD}{#2}
    \put(\value{robA},\value{robD}){$c$}
}

\def\trojkatLevelDwa(#1,#2,#3){
\trojkat(#1,#2,#3)
\setcounter{robF}{#1}%
\addtocounter{robF}{#3}%
\trojkat(\value{robF},#2,#3)
\setcounter{robE}{#3}
\divide\c@robE2
\addtocounter{robE}{#1}
\setcounter{robG}{#3}
\multiply\c@robG3
\divide\c@robG4
\addtocounter{robG}{#2}
\trojkat(\value{robE},\value{robG},#3)
}
\title[Motif based hierarchical random graph]%
{Motif based hierarchical random graphs: structural properties and
critical points of an Ising model\footnote{This work was supported
by the DFG through the project 436 POL 125/0--1 as well as through
SFB 701: ``Spektrale Strukturen und Topologische Methoden in der
Mathematik''. Yuri Kozitsky was also supported by TODEQ
MTKD--CT--2005--030042.} }
\author[M.~Kotorowicz, Yu.~Kozitsky]{Monika Kotorowicz\thanks{E-mail:
monika@hektor.umcs.lublin.pl}\,, Yuri Kozitsky\thanks{E-mail:
jkozi@hektor.umcs.lublin.pl}}
\address{Institute of Mathematics, Maria Curie-Sk{\l}odowska University,
Lublin, Poland}

\authorcopyright{M.~Kotorowicz, Yu.~Kozitsky}
\date{Received March 15, 2010, in final form June 7, 2010}

\begin{document}

\maketitle

\begin{abstract}
A class of random graphs is introduced and studied. The graphs are
constructed in an algorithmic way from five motifs which were
found in [Milo~R., Shen-Orr~S., Itzkovitz~S., Kashtan~N.,
Chklovskii~D., Alon~U., Science, 2002, \textbf{298}, 824--827].
The construction scheme resembles that used in [Hinczewski~M.,
A.~Nihat Berker, Phys. Rev.~E, 2006,  \textbf{73}, 066126],
according to which the short-range bonds are non-random, whereas
the long-range bonds appear independently with the same
probability. A number of structural properties of the graphs have
been described, among which there are degree distributions, clustering,
amenability, small-world property. For one of the motifs, the
critical point of the Ising model defined on the corresponding
graph has been studied.

\pacs 89.75.Fb, 89.75.Kd, 05.10.Cc, 05.70.Jk
\end{abstract}

\section{Introduction and setup}

A vast variety of large systems occurring in nature and society
have a very  complicated topological structure. These are the
Internet, the World Wide Web, citation, neural, and  social
networks, etc. In view of the complex topology and unknown
organizing principles, the networks are often modeled as random
graphs. A random graph with a given node set $V$ is a graph in
which for a given pair $i,j\in V$, the bond $\langle i,j \rangle$
appears at random. The study of random graphs has been originated
by P.~Erd\H{o}s and A.~R\'{e}nyi, who were the first to introduce
such graphs in~\cite{ErdosRenyi1959,ErdosRenyi1960}. In the
Erd\H{o}s-R\'{e}nyi random graph model,  denoted by~$G_{n,p}$\,,
the number of nodes is $n$, and the bonds between distinct nodes
appear independently\footnote{C.f., however, the discussion
in~\cite{Burda1}.} with the same probability~$p$. An important
characteristic of a graph is the {\it node degree}, which is the
number of bonds attached thereat. In $G_{n,p}$\,, it is a random
variable, and all such variables are independent and have the same
probability distribution. Namely, the probability that the degree
of a given node is $k$ is given by the Bernoulli law
\begin{equation} \label{i}
  P_n (k) = {n-1 \choose k}p^k(1-p)^{n-1-k}\,, \qquad k=0, 1 , \dots, n-1.
\end{equation}
For random graphs, the most interesting questions refer to their
asymptotic properties in the limit $n\rightarrow + \infty$. To get
nontrivial answers to such questions one allows the parameters
to depend on $n$. For $p_n = c/n$, the limit of (\ref{i}) is the
Poisson law
\begin{equation}
 \label{j}
P(k) = {c^k} e^{-c}/k!\,, \qquad k\in \mathbb{N}_0.
\end{equation}
However, for irregular complex networks, the~random graph
$G_{n,p}$ is~not a~good model since in the~most of such networks
the degree distribution is essentially non-Poissonian. Many real
world networks, e.g. the WWW, are characterized by  power-law node
degree distributions, which have the form $P(k) = Ck^{-\gamma}$,
$\gamma
>1$, typical for the so-called {\it scale-free} graphs.

Another important parameter characterizing a random graph is the
{\it clustering coefficient}, which is the probability that two
nodes are neighbors given they have a common neighbor. Clearly, in
$G_{n,p}$ this probability is $p$, and is the same independently
of whether or not the nodes have a common neighbor. Real world
networks usually manifest strong clustering, which once more
indicates that Erd\H{o}s-R\'{e}nyi type random graphs are not
appropriate as their models. In~\cite{WattsStrogatz1998},
D.J.~Watts and H.~Strogatz proposed another type of~random graphs,
in which the mentioned disadvantage is overcome. It should be
noted, however, that such graphs do not have power law node degree
distributions. The  next step beyond the Erd\H{o}s-R\'{e}nyi
model was done by A.-L.~Barab\'{a}si and R.~Albert
in~\cite{BarabasiAlbert1999}. In their model, the preferential
attachment principle has been employed, typical for many real
networks (the Internet, citation and social networks). According
to this principle, the more connected a node is, the more likely
it  receives a new bond. The construction of the
Barab\'{a}si-Albert model starts from an initial graph with $m_0
\geqslant 2$ nodes (neither can be isolated). At each step, one
adds a new node and connects it to the existing nodes. The
probability $p_i$ that the new node is connected to node $i$ is
proportional to the degree $n_i$ of that node, that is, $p_i =
n_i/(\sum_j n_j)$.

\begin{figure}[htbp]
\centering
\begin{picture}(300,45)
\trojkat(0,0,50)
\trojkat(70,0,50)
\put(95,38){\line(1,0){25}}
\put(140,0){\framebox(37,37){}}
\put(200,0){\framebox(37,37){}}
\put(260,0){\framebox(37,37){}}
\put(200,0){\line(1,1){37}}
\put(260,0){\line(1,1){37}}
\put(297,0){\line(-1,1){37}}
\end{picture}
\caption{Three and four node motifs $M_1$, $M_2$, $M_3$, $M_4$, $M_5$
found in~\cite{ItzkovitzAlon}.} \label{motifs}
\end{figure}
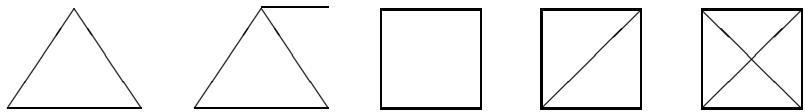

Many of the complex networks occurring in nature contain
characteristic patterns, recurring much more frequently than the
other ones. They are called {\it network motifs},
see~\cite{Milo,ItzkovitzMiloKashtan2003,Alon,Kashtan,ItzkovitzAlon,Matias}.
Different networks may have different motifs, and motifs in turn
can characterize the networks. For instance, in biological
regulation networks it has been experimentally demonstrated that
each of the motifs can perform a key information processing
function, see~\cite{Alon}. In~\cite{ItzkovitzAlon}, the authors
introduced a random graph model based on some geometric principles
(constraints). Then, they compared the appearance of eight
elementary three- and four-node patterns in their model with the
same characteristics of the Erd\H{o}s-R\'{e}nyi random graph. It
turned out that five of these patterns are motifs for their model,
but not for the Erd\H{o}s-R\'{e}nyi random graph, see
figure~\ref{motifs} below and table~1 in~\cite{ItzkovitzAlon}.
For another random graph model, the appearance of the same eight
patterns was also studied in~\cite{Matias}. Note that among these
patterns, only $M_1$ and $M_5$ correspond to complete graphs (each
node is a neighbor to every other node).

One of the ways to get information about infinite graphs, also
random ones, is to study the properties of certain models of
statistical physics defined thereon. The most popular ones are the
Ising and Potts models, as well as the models of bond and site
percolation, see~\cite{Hag}. On the other hand, in statistical
physics certain graphs are employed  to mimic a crystal lattice,
for which the critical-point behavior of the Ising model can be
described in an explicit and rigorous way. These are the so-called
{\it hierarchical lattices} introduced
in~\cite{GriffithsKaufman1982}. Such lattices are defined in a
rather algorithmic way by means of the basic pattern, e.g., by a
``diamond'', which is the pattern $M_3$ depicted in
figure~\ref{motifs}. A mathematical description of the Gibbs
states of the Ising model on such graphs was done by P.M.~Bleher
and E.~\v{Z}alys in~\cite{BZ,BleherZalys1989}.  M.~Hinczewski and
A.~Nihat Berker~\cite{HinczBerker2006} studied the critical-point
properties of the Ising model on the diamond hierarchical lattice
`decorated' by additional bonds, which appear at random. In the
present paper, we follow the way suggested
in~\cite{HinczBerker2006} and introduce hierarchical graphs
constructed by means of the motifs shown in figure~\ref{motifs},
decorated by additional bonds which appear at random and repeat,
in a way, the corresponding motif. We analyse some of their
characteristics, such as the average degree, the node degree
distribution, amenability, the small-world property, as well as
the critical-point properties of the Ising model. This study of ours
is a continuation of~\cite{Wrobel}, where the graph based on $M_1$
was introduced by one of us. Note that our hierarchical graphs also
found applications in cryptography, see~\cite{Kotorowicz}.

\section{The graphs: construction and structural properties}

\subsection{The construction: informal description}

As is typical for hierarchical graphs, e.g., for hierarchical
lattices in~\cite{GriffithsKaufman1982,HinczBerker2006}, the
construction is carried out in an algorithmic way: at $k$-th
level, $k\in \mathbb{N}$, one produces a subgraph, say
$\Lambda_k$\,, which is  then used as a construction element for
producing $\Lambda_{k+1}$\,. The procedure is the same at each
level, and the starting element is obtained from the corresponding
motif. Let us illustrate this for the simplest case based on the
motif $M_1$\,. First we label the nodes of $M_1$ by $a$, $b$, and
$c$, as it is shown in figure~\ref{construction}, and obtain
$\Lambda_1$~--- the triangle. Then we take three such graphs and
label them by the same labels. As a result, each of the
triangles has nodes of two kinds: one node the label of which
coincides with the triangle label (e.g., node $a$ in triangle $a$),
and two nodes with non-coinciding labels. Thereafter, the
triangles are glued up according to the following rule: node
$c$ of triangle $a$ is glued up with node $a$ of triangle $c$,
ect. The nodes with the coinciding labels remain untouched. These
are the so-called {\it external} nodes of $\Lambda_2$\,. The
remaining nodes are called {\it internal}. The bonds of the
initial triangles turn into the bonds of $\Lambda_2$\,. We call
them {\it basic} bonds; they are depicted by solid lines. At the
next stage of the first step, we add the bonds connecting the external
nodes in the same way as it is in the motif $M_1$\,. Such bonds
are depicted by dotted lines and are called {\it decorations}. As a
result, we obtain the graph $\Lambda_2$\,, which has six basic
bonds and three decorations, three external and three internal
nodes. Then we repeat the same procedure~--- take three copies of
$\Lambda_2$\,, label them by $a$, $b$, and $c$, and divide their
external nine nodes into two groups: three nodes with coinciding
labels and six nodes with non-coinciding labels. Then the graphs
$\Lambda_2$ are glued up as described above. Thereafter, three
decorating bonds are drawn to connect the external nodes. This
procedure is repeated ad infinitum. Similar constructions for
motifs $M_2$ and $M_3$ are presented in figure~\ref{motif2} and
figure~\ref{4444}, respectively. Here we have omitted the
decorating bonds  not to overload the pictures. The
picture for $M_4$ is obtained from that for $M_3$ by adding the
diagonals. The picture for $M_5$ is just the three-dimensional
version of the picture for $M_1$\,, where the basic pattern is a
tetrahedron.

\subsection{Definitions}

In order to fix the terminology and to make the construction of
our graphs mathematically immaculate, we begin by introducing a
number of relevant mathematical notions. A (simple) graph ${\sf
G}$ is a pair of sets $({\sf V}, {\sf E})$, where ${\sf V}$ is the
set of nodes, whereas ${\sf E}$ is a subset of the Cartesian
product ${\sf V} \times {\sf V}$. We suppose that ${\sf E}$ is
symmetric and irreflexive, i.e., $\langle j, i \rangle \in {\sf
E}$ whenever $\langle  i ,j\rangle \in {\sf E}$, and  $\langle i,
i \rangle \notin {\sf E}$ for every $i, j\in {\sf V}$. We say that
$i$ and $j$ are connected by a {\it bond} if $\langle i, j \rangle
\in {\sf E}$. In this case, we write $i \sim j$ and say that $i$
and $j$ are {\it adjacent} or that they are {\it neighbors}.
Hence, the elements of ${\sf E}$ themselves can be called bonds.
The graph is said to be {\it complete} if each two nodes are
adjacent. For a given $i$, by $n(i)$ we denote the {\it degree} of
$i$~--- the number of its neighbors. If ${\sf V}$ is finite, then
${\sf E}$ is also finite, and the graph is said to be finite.
Otherwise, the graph is infinite. An infinite graph is called {\it
locally finite}, if $n(i)$ is finite for every node. All the
infinite graphs we study are {\it countable}, which means that
both sets ${\sf V}$ and ${\sf E}$ are infinite and countable.
Given ${\sf G}=({\sf V}, {\sf E})$ and ${\sf G}'=({\sf V}', {\sf
E}')$, let $\phi: {\sf V} \rightarrow {\sf V}'$ be such that $\phi
(i) \sim \phi(j)$ whenever $i\sim j$. Such a map $\phi$ is called
a {\it morphism}. A bijective morphism is called an {\it
isomorphism}. If $\phi$ is an isomorphism, then its inverse
$\phi^{-1}$ is also an isomorphism, and then the graphs ${\sf G}$
and ${\sf G}'$ are said to be mutually isomorphic. Such graphs
have identical structures and thus can be identified. In this
case, we also say that ${\sf G}'$ is a {\it copy} of ${\sf G}$.
One observes that this refers to both finite and infinite graphs.
An isomorphism $\phi : {\sf V} \rightarrow {\sf V}$, i.e. which
maps the graph onto itself, is called an {\it automorphism}. The
graph ${\sf G}'=({\sf V}', {\sf E}')$ such that ${\sf V}' \subset
{\sf V}$ and ${\sf E}' \subset {\sf E}$ is said to be a {\it
subgraph} of ${\sf G}=({\sf V}, {\sf E})$. In this case, we write
${\sf G}' \subset {\sf G}$. Suppose that a subgraph ${\sf G}'
\subset {\sf G}$ has a copy, say $\widetilde{\sf G}$, that is,
there exists an isomorphism $\phi : \widetilde{\sf G} \rightarrow
{\sf G}'$. Then $\phi$, considered as a map $\phi : \widetilde{\sf
G} \rightarrow {\sf G}$, is called an {\it embedding} of
$\widetilde{\sf G}$ into ${\sf G}$, whereas ${\sf G}'$ is called
the {\it image} of $\widetilde{\sf G}$ under this embedding.
Figure~\ref{motifs} presents the so-called {\it unlabeled} graphs,
which we call patterns. After labeling, i.e., attaching a label to
each of the nodes, such a pattern turns into a graph. Another
labeling may or may not yield the same graph. This depends on
whether or not there exists the corresponding automorphism. For
instance, any labeling of the triangle $M_1$ yields the same graph
since in any case each of the nodes has the same neighbors. For the
pattern $M_2$\,, the left-hand graph in figure~\ref{motif2} with
the interchanged labels $a$ and $b$ is the same. However, the
graph with the interchanges $c$ and $d$ is not the same anymore.
Of course, this new graph is isomorphic to the initial one. This
is because there is only one nontrivial automorphism of $M_2$\,:
the one which interchanges $a$ and $b$, and preserves $c$ and $d$.
The triangle has six automorphisms.

Let us now turn to {\it random} graphs. To introduce such a graph
we need an {\it underlying} graph ${\sf G}=({\sf V}, {\sf E})$ and
a family $\mathcal{E}$ of subsets of ${\sf E}$. If ${\sf G}$ is
finite, as $\mathcal{E}$ one can take the set of all subsets of
${\sf E}$. In the sequel, we deal with such graphs only. Thus, for
${\sf E}'\in \mathcal{E}$, we say that ${\sf E}'$ has been picked
{\it at random} with probability $P({\sf E}')$. In  the
Erd\H{o}s-R\'enyi model $G_{n,p}$\,, the underlying graph is
complete with the node set ${\sf V} = \{1 , \dots , n\}$. In this
model, $P({\sf E}') = p^{|{\sf E}'|}$, where $p\in [0,1]$ and
$|{\sf E}'|$ stands for the number of elements in ${\sf E}'$. In
other words, the elements of ${\sf E}$ are being picked
independently, each  with the same probability $p$. In a bit
complicated model, the bonds are picked independently but with
probability which depends on the bond. In this case, as well as in
the case of the Erd\H{o}s-R\'enyi model,  we deal with a random
graph with independent bonds. For such graphs,
\begin{equation}
  \label{gra}
  P({\sf E}') = \prod_{e\in {\sf E}'} p(e),
\end{equation}
where $p(e)$ is the probability of picking bond $e$. The set of
graphs ${\sf G}' = ({\sf V}, {\sf E}')$ with ${\sf E}'\in
\mathcal{E}$ is called the graph {\it ensemble}~--- each ${\sf
G}'$ is picked at random from this ensemble. A {\it random
graph model} is the pair consisting of the graph ensemble and
of the function $P:\mathcal{E}\rightarrow [0,1]$. If the function
$P$ is as in~(\ref{gra}), the graph is said to be a random graph
with independent bonds. Suppose that we have two random graph
models with independent bonds. Let also ${\sf G}=({\sf V}, {\sf
E})$ and $\widetilde{\sf G}=(\widetilde{\sf V}, \widetilde{\sf
E})$ be their underlying graphs and $\phi: {\sf V} \rightarrow
\widetilde{\sf V}$ be a morphism. Then this map is said to be the
{\it morphism of the random graphs} if for every $\langle i, j
\rangle \in {\sf E}$, the probability (in the first model) that
this bond is picked is the same as the corresponding
probability (in the second model) for the bond $\langle \phi(i),
\phi(j)\rangle$.

\subsection{The construction}

As was mentioned above, each of our graphs is constructed in an
algorithmic way from the corresponding motif presented in
figure~\ref{motifs}. Since they are going to be  random graphs with
independent bonds,  we have to construct the corresponding
underlying graphs and, to define, for a given bond, the probability
of being picked, c.f.~(\ref{gra}). In all our models, the bonds
will be of two kinds, which we call basic bonds and decorations.
Basic bonds are going to be non-random, i.e. they are picked with
probability one. Decorating bonds appear with probability $p\in
[0,1]$, which is a parameter of the model. Turn now to the
construction of the underlying graphs. Let $q$ be the number of
nodes in the corresponding motif, that is, $q=3$ for $M_1$ and
$q=4$ for the remaining motifs. At step $k=1$, we just label the
vertices of the corresponding motif by $i=1, \dots , q$ and obtain
the initial graph $\Lambda_1 = (V_1\,, E_1)$. All its bonds are
set to be basic.  Suppose now that we have $q+1$ copies of
$\Lambda_1$ obtained by the isomorphisms $\phi_2^j$\,, $j= 0,1,
\dots , q$. Thus, in $j$-th copy the nodes are $\phi^j_2(i)$,
$i=1, \dots , q$. The graph $\Lambda_2$ is obtained from these
copies under the following conditions
\begin{equation}
  \label{gra2}
 \phi_2^0 (i) = \phi_2^i (i), \qquad i=1, \dots , q;
 \qquad \phi_2^i(j) = \phi_2^j (i), \qquad i=1, \dots , q, \ i\neq j.
\end{equation}
Thus, the images of $V_1$ under $\phi_2^i$ and $\phi_2^j$ with
$i\neq j$ intersect only at one node where~(\ref{gra2}) holds. The
maps $\phi_2^j$\,, $j= 0,1, \dots , q$ embed $\Lambda_1$ into
$\Lambda_2$\,. The nodes $\phi_2^i(i)$, $i=1, \dots , q$, are
called the {\it external} nodes of $\Lambda_2$\,. All other nodes
are called {\it internal}. Thus, $\Lambda_2$ has $q$ external and
$q(q-1)/2$ internal nodes. At this stage, we label them by $i= 1 ,
\dots , q(q+1)/2$ in such a way that the external nodes have the
same labels as in $\Lambda_1$\,, that is,  $\phi_2^i(i) = i$,
$i=1, \dots q$. By construction, the bonds obtained as images
under the map $\phi_2^0$ are decorations: they are of the form
$\langle \phi^0_2 (i), \phi^0_2 (j) \rangle$ where $i$ and $j$ are
adjacent in $\Lambda_1$\,. From the first condition
in~(\ref{gra2}) we see that the decorating bonds connect the
external nodes of $\Lambda_2$\,. The remaining bonds of
$\Lambda_2$ are set to be basic. Now we construct $\Lambda_3$ from
one copy of $\Lambda_1$ and $q$ copies of $\Lambda_2$\,. Let
$\phi_3^0$ be the map which produces the copy of $\Lambda_1$ and
$\phi_3^j$\,, $j=1, \dots ,q$ be the maps which produce the copies
of $\Lambda_2$\,. We then impose the conditions
\begin{equation}
  \label{gra3}
  \phi_3^0 (i) = \phi_3^i (i), \qquad i=1, \dots , q;
  \qquad \phi_3^i(j) = \phi_3^j (i), \qquad i=1, \dots , q, \ i\neq j
\end{equation}
and obtain $\Lambda_3$. Thus,  $\phi_3^0$ embeds $\Lambda_1
\rightarrow \Lambda_3$\,, and $\phi_3^i :\Lambda_2 \rightarrow
\Lambda_3$\,, $i=1, 2,\dots, q$. As above, the nodes $\phi_3^i
(i)$ are set to be external, and the remaining nodes are internal.
The images of $V_2$ under $\phi_3^i$ and $\phi_3^j$ with $i\neq j$
intersect only at one node where~(\ref{gra3}) holds. Again we
label the nodes of $\Lambda_3$ in such a way that $\phi_3^i(i) =
i$, $i=1, \dots , q$. Now let us establish which bonds of
$\Lambda_2$ are decorating and which are basic. As above, the
bonds connecting the external nodes are decorating. The images of
decorating bonds of $\Lambda_2$ are decorating bonds in
$\Lambda_3$\,; the same is  also true for the basic bonds~--- the
basic bonds of $\Lambda_3$ are exactly the images of the basic
bonds of $\Lambda_2$\,. For $k\geqslant 4$, the construction of
$\Lambda_{k}$ from $\Lambda_{k-1}$ and $\Lambda_1$ is identical to
the construction of $\Lambda_3$ just described. As above, by $V_k$
and $E_k$ we denote the sets of nodes and bonds of $\Lambda_k$\,,
respectively. Thus, for $k\geqslant 2$ we have $E_k = E_k' \cup
E_k''$\,, where $E_k'$ (respectively, $E_k''$) consists of basic
(respectively, decorating) bonds. All $\Lambda_k$\,, $k\in
\mathbb{N}$, are considered as subgraphs of an infinite graph
$\Lambda_\infty$\,, the structure and properties of which are not
important for the study presented in this article.

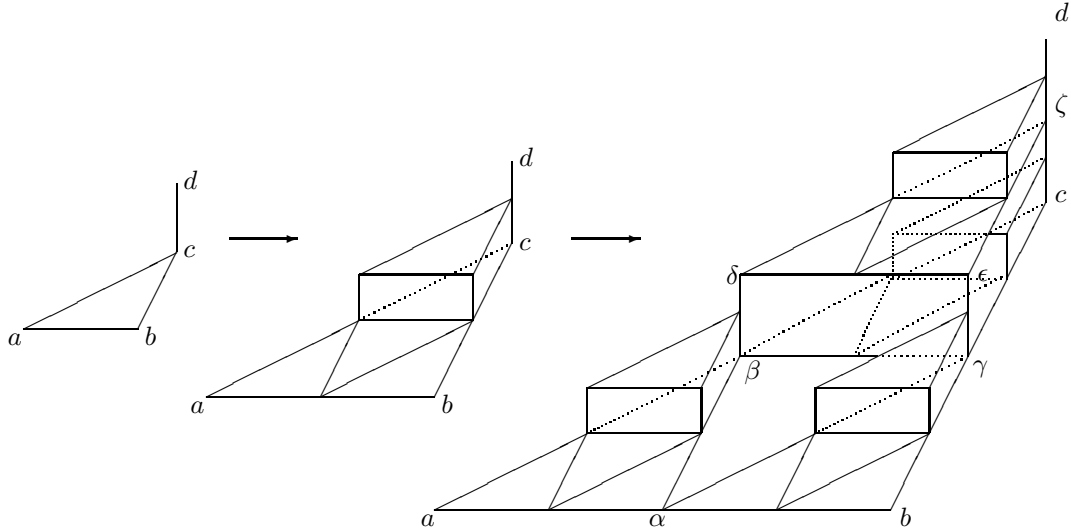
\begin{figure}[ht]
\centering
\begin{picture}(400, 200)
\unitlength 0,3mm
    \put(0,90){\line(1,0){50}}
    \put(0,90){\line(2,1){67}}
    \put(50,90){\line(1,2){17}}
    \put(67,124){\line(0,1){30}}

    \put(-7,83){$a$}
    \put(53,83){$b$}
    \put(70,122){$c$}
    \put(70,152){$d$}

\put(90,130){\vector(1,0){30}}

    \put(80,60){\line(1,0){50}}
    \put(80,60){\line(2,1){67}}
    \put(130,60){\line(1,2){17}}
    \put(147,94){\line(0,1){20}}

    \put(130,60){\line(1,0){50}}
    \put(130,60){\line(2,1){67}}
    \put(180,60){\line(1,2){17}}
    \put(197,94){\line(0,1){20}}

    \put(147,94){\line(1,0){50}}
    \qbezier[30](147,94)(180,110)(214,128)
    \put(197,94){\line(1,2){17}}
    \put(214,128){\line(0,1){20}}

    \put(147,114){\line(1,0){50}}
    \put(147,114){\line(2,1){67}}
    \put(197,114){\line(1,2){17}}
    \put(214,144){\line(0,1){20}}

    \put(73,53){$a$}
    \put(183,53){$b$}
    \put(217,125){$c$}
    \put(217,162){$d$}

\put(240,130){\vector(1,0){30}}

\put(174,3){$a$}

    \put(180,10){\line(1,0){50}}
    \put(180,10){\line(2,1){67}}
    \put(230,10){\line(1,2){17}}
    \put(247,44){\line(0,1){20}}

    \put(230,10){\line(1,0){50}}
    \put(230,10){\line(2,1){67}}
    \put(280,10){\line(1,2){17}}
    \put(297,44){\line(0,1){20}}

    \put(247,44){\line(1,0){50}}
    \qbezier[30](247,44)(270,55)(314,78)
    \put(297,44){\line(1,2){17}}
    \put(314,78){\line(0,1){20}}

    \put(247,64){\line(1,0){50}}
    \put(247,64){\line(2,1){67}}
    \put(297,64){\line(1,2){17}}
    \put(314,94){\line(0,1){20}}


\put(274,3){$\alpha$}
\put(384,3){$b$}
    \put(280,10){\line(1,0){50}}
    \put(280,10){\line(2,1){67}}
    \put(330,10){\line(1,2){17}}
    \put(347,44){\line(0,1){20}}

    \put(330,10){\line(1,0){50}}
    \put(330,10){\line(2,1){67}}
    \put(380,10){\line(1,2){17}}
    \put(397,44){\line(0,1){20}}

    \put(347,44){\line(1,0){50}}
    \qbezier[30](347,44)(370,55)(414,78)
    \put(397,44){\line(1,2){17}}
    \put(414,78){\line(0,1){20}}

    \put(347,64){\line(1,0){50}}
    \put(347,64){\line(2,1){67}}
    \put(397,64){\line(1,2){17}}
    \put(414,94){\line(0,1){20}}

\put(316,68){$\beta$}
\put(416,71){$\gamma$}
\put(452, 146){$c$}
    \put(314,78){\line(1,0){50}}
    \qbezier[30](314,78)(345,96)(381,114)
    \qbezier[20](364,78)(372,96)(381,114)
    \put(381,112){\line(0,1){2}}
    \qbezier[10](381,114)(381,123)(381,132)

    \put(364,78){\line(1,0){10}}
    \qbezier[15](374,78)(394,78)(414,78)
    \qbezier[30](364,78)(395,96)(431,114)
    \put(414,78){\line(1,2){17}}
    \put(431,112){\line(0,1){20}}

    \qbezier[20](381,112)(404,112)(431,112)
    \qbezier[30](381,112)(420,130)(448,146)
    \put(431,112){\line(1,2){17}}
    \put(448,146){\line(0,1){20}}

    \put(423,132){\line(1,0){8}}
    \qbezier[20](381,132)(402,132)(423,132)
    \qbezier[35](381,132)(414,148)(446,165)
    \put(431,132){\line(1,2){17}}
    \put(448,162){\line(0,1){20}}

    \put(308,110){$\delta$}
    \put(418,111){$\epsilon$}
    \put(452,186){$\zeta$}
    \put(452, 226){$d$}
    \put(314,114){\line(1,0){50}}
    \put(314,114){\line(2,1){67}}
    \put(364,114){\line(1,2){17}}
    \put(381,148){\line(0,1){20}}

    \put(364,114){\line(1,0){50}}
    \put(364,114){\line(2,1){67}}
    \put(414,114){\line(1,2){17}}
    \put(431,148){\line(0,1){20}}

    \put(381,148){\line(1,0){50}}
    \qbezier[30](381,148)(420,166)(448,182)
    \put(431,148){\line(1,2){17}}
    \put(448,182){\line(0,1){20}}

    \put(381,168){\line(1,0){50}}
    \put(381,168){\line(2,1){67}}
    \put(431,168){\line(1,2){17}}
    \put(448,198){\line(0,1){20}}

\end{picture}
\caption{Construction of the bare graph $\Lambda_3$ based on
$M_2$\,.} \label{motif2}
\end{figure}

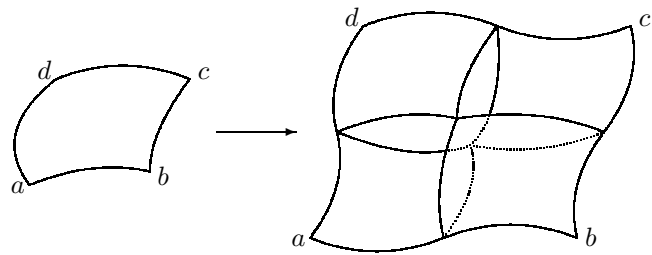
\begin{figure}[htbp] 
\centering
\begin{picture}(300, 120)
\qbezier[150](50,70)(75,80)(100,70)
\qbezier[150](40,30)(65,40)(85,35)
\qbezier[150](50,70)(25,50)(40,30)
\qbezier[150](100,70)(85,50)(85,35)
    \put(43,70){$d$}
    \put(103,70){$c$}
    \put(33,27){$a$}
    \put(88,30){$b$}
\put(110,50){\vector(1,0){30}}
\qbezier[150](165,90)(190,100)(215,90)
\qbezier[150](155,50)(180,60)(200,55)
\qbezier[150](165,90)(150,70)(155,50)
\qbezier[150](215,90)(200,70)(200,55)
\qbezier[150](200,55)(230,60)(255,50)
\qbezier[150](195,10)(220,20)(245,10)
\qbezier[150](200,55)(190,30)(195,10)
\qbezier[150](255,50)(240,30)(245,10)
\qbezier[150](215,90)(240,80)(265,90)
\qbezier[40](205,45)(230,40)(255,50)
\qbezier[100](215,90)(217,70)(212,56)
\qbezier[10](212,56)(210,50)(205,45)
\qbezier[150](265,90)(270,70)(255,50)
\qbezier[100](155,50)(180,40)(196,43)
\qbezier[8](196,43)(200,43)(205,45)
\qbezier[150](145,10)(170,0)(195,10)
\qbezier[150](155,50)(160,30)(145,10)
\qbezier[30](205,45)(210,30)(195,10)
    \put(158,90){$d$}
    \put(268,90){$c$}
    \put(138,7){$a$}
    \put(248,7){$b$}
\end{picture}
\caption{Construction of the bare graph $\Lambda_2$ based on
$M_3$\,. }\label{4444}
\end{figure}

Note that the construction principle used above  essentially
differs from that used in the construction of the hierarchical
lattices
in~\cite{GriffithsKaufman1982,BZ,BleherZalys1989,HinczBerker2006}.
Namely, in our case to obtain $\Lambda_k$ one replaces each {\it
node} of the basic pattern by a copy of the graph
$\Lambda_{k-1}$\,. In the hierarchical lattices, one replaces a
{\it bond}. As we shall see in the sequel, this leads to
essentially different properties of the resulting graphs. Below in
figure~\ref{construction}, we illustrate the construction
described above for the case where the basic pattern is the motif
$M_1$\,. In this case, the bare graph (i.e. the one which occurs
for $p=0$) is the approximating graph for the fractal known as the
Sierpi\'nski gasket\footnote{The fractal itself is obtained as the
closure of the set $\cup_{k\in \mathbb{N}} V_k$ in the appropriate
topology, see e.g.~\cite{Barlow}.}. The elements of $E'_2$ (middle
graph) and of $E'_3$ (right-hand graph) are depicted by solid
lines. The elements of $E''_2$  and of $E''_3$ are depicted  by
dotted lines. We omit some dotted lines to indicate that they
appear at random and hence may be absent in a given realization of
the graph. Note that $\Lambda_3$ can be viewed as the triangle
composed of three copies of $\Lambda_2$\,. In
figure~\ref{motif2}, we present the construction of the bare graph
$\Lambda_3$ corresponding to $M_2$\,. In contrast to the former
case, this is not a planar graph. In figure~\ref{4444}, we construct
the bare graph $\Lambda_2$ for the motif $M_3$\,. One observes
that in that picture the node $c$ of the lower left-hand quadrat
(i.e. quadrat $a$) is glued up with node $a$ of the upper
right-hand quadrat. It is interesting that the corresponding
fractal can be obtained by the following procedure, resembling the
one which yields the Sierpi\'nski gasket. One takes the full
quadrat and cuts it out into four equal quadrats, but without
cutting the external lines. Then, one glues up the vertices of the
smaller quadrats as depicted and proceeds with cutting out the
smaller quadrats. The fractal which one obtains from $M_5$ is a
three dimensional version of the Sierpi\'nski gasket. One takes
the full tetrahedron and cuts out its inner one fourth in such a
way that the remaining four tetrahedra are glued up according to
the rule: vertex $b$ of tetrahedron $a$ is glued up with vertex $a$
of tetrahedron $b$, etc.
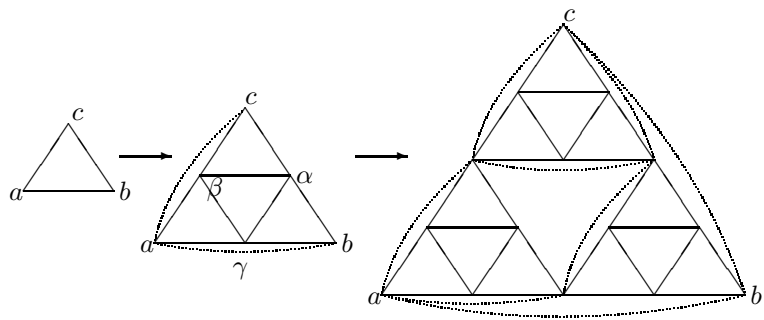
\begin{figure}[htbp]
\centering
\unitlength 0,23mm
\begin{picture}(400, 180)
\trojkatABC(10,70,52)
\put(65,90){\vector(1,0){30}}
\trojkatLevelDwa(85,40,52)
\put(77,35){$a$}
\put(192,35){$b$}
\put(130,22){$\gamma$}
\put(137,120){$c$}
\put(115,67){$\beta$}
\put(166,75){$\alpha$}
\qbezier[50](85,40)(93,80)(135,116)
\qbezier[50](85,40)(135,30)(189,40)
\put(200,90){\vector(1,0){30}}
\trojkatLevelDwa(215,10,52)
\trojkatLevelDwa(319,10,52)
\trojkatLevelDwa(267,88,52)
\put(207,5){$a$}
\put(426,5){$b$}
\put(319,168){$c$}
\qbezier[50](215,10)(223,50)(267,88)
\qbezier[50](215,10)(265,0)(319,10)
\qbezier[50](319,10)(325,50)(371,88)
\qbezier[50](267,88)(273,126)(319,166)
\qbezier[50](371,88)(360,126)(319,166)
\qbezier[50](267,88)(315,76)(371,88)
\qbezier[100](215,10)(315,-15)(423,10)
\qbezier[100](423,10)(395,95)(319,166)
\end{picture}
\caption{Construction of the graph $\Lambda_3$ based on $M_1$\,. }
\label{construction}
\end{figure}

\subsection{Degree distribution}

Now we turn to the description of the structural properties of the
graphs constructed above. Let $m_k$\,, $k\in \mathbb{N}$, be the
number of times the basic pattern appears in non-decorated $\Lambda_k$
as a subgraph. For the graphs based on $M_1$ and $M_5$ we have the same situation. Here $m_1=1$ and $m_k=qm_{k-1}\,$, for
$k\geqslant2$ with the exception in $\Lambda_2\,$, where the additional pattern appears. So, for $M_1$ and $M_5$
we obtain $m_1=1$ and
\begin{equation}\label{gra10}
m_k = (q+1)q^{k-2}, \qquad k\geqslant
2.
\end{equation}
Here $q=3$ and $q=4$ for $M_1$ and $M_5\,$, respectively. For motif $M_3$ we have $m_1=1$, $m_2=2q$ and
\begin{equation}
  \label{MM5}
 m_k = qm_{k-1}, \qquad k\geqslant 3.
\end{equation}
Hence $m_k=2\cdot4^{k-1}$ for $k\geqslant2$. The simplest case if for $M_4\,$, where $m_k=qm_{k-1}$ for $k\geqslant2$. It gives $m_4=4^{k-1}$. The last motif is $M_2$~-- triangle with additional bond. On each level this bond ``produces'' new 17 patterns. So
$m_1=1$ and for $k\geqslant2$ \ $m_k=\frac{1}{3}(26\cdot4^{k-1}-17)$.

Now we analyze the number of times the basic pattern appears in fully decorated $\Lambda_k\,$, denoted by $\widetilde{m}_k\,$. For the graphs based on $M_1$ and $M_5$ we have
\begin{equation}
 \label{m1}
\widetilde{m}_k = \frac{2q+1}{q-1}q^{k-1}+\frac{q+2}{q-1}\,.
\end{equation}
For $M_3$ it is $\frac{2}{3}4^{k}-\frac{5}{3}$ and for $M_4$ we obtain $\frac{1}{3}(4^k-1)$.
In all cases, we have $m_k$ increasing as $C(p) q^{k-1}$, which means that adding decorations does not
change the asymptotics of $m_k\,$.

In a similar way, we obtain
\begin{equation}
 \label{m2}
|V_k| = \frac{q^{k} +q}{2}\,, \qquad  |E_k| = r q^{k-1} + r p
\frac{q^{k-1}-1}{q-1}\,, \qquad k \in \mathbb{N},
\end{equation}
where $|V_k|$ stands for the number of nodes in $\Lambda_k$\,,
whereas $|E_k|$ is the expected number of bonds in this graph.

As was mentioned above, the degree distribution is a very
important characteristic of the graph. In contrast to
Erd\H{o}s-R\'enyi type graphs, for our graphs the distribution of
the random variable $n(i)$ depends on the type of $i$. Thus, the
simplest way to describe this distribution is to average $n(i)$
over the nodes of a given $\Lambda_k$\,, that is, to consider
 \begin{equation}
   \label{Q1}
  n_k = \frac{1}{|V_k|} \sum_{i \in V_k} n(i).
 \end{equation}
 Let $\langle n_k \rangle$ be the expected value of $n_k$\,. Then
\begin{equation}
 \label{n3}
\langle n_k \rangle = 2 |E_k|/ |V_k| = \frac{4 r}{q(q-1)}( q-1+p)-
\frac{4 r}{q(q-1)} \cdot \frac{q-1 + 2p}{ q^{k-1}+1}\,.
\end{equation}
However, this result provides only partial information on the node
degree distribution. To get more, let us analyze the structure of
the node sets $V_k$\,, $k=1 , 2 , \dots  $, more in detail. For a
given $\Lambda_k$ and $l = 1 , \dots , k$, let $V^{(l)}_k$ be the
set of nodes $i\in V_k$ which are external for some $\Lambda_l$
and, at the same time, are internal for any $\Lambda_{l+1}$\,. Of
course, here we mean those $\Lambda_l$'s which are subgraphs for
$\Lambda_k$\,. As an example, let us consider the graph
$\Lambda_2$ based on $M_1$\,, see the middle graph in
figure~\ref{construction}. The nodes $a$, $b$, and $c$ constitute
$V^{(2)}_2$, whereas the remaining nodes constitute $V^{(1)}_2$.

The elements  of $V^{(k-1)}_k$ are exactly the nodes at which the
subgraphs $\Lambda_{k-1}^j$\,, $j=1, \dots , q$ are glued up to
form $\Lambda_k$\,, whereas the elements of $V^{(k)}_k$ are
exactly the external nodes of $\Lambda_k$\,. Then $|V^{(k)}_k|=q$
and $|V^{(k-1)}_k| = q (q-1)/2$. For $l< k-1$, we have
$|V^{(l)}_{k}| = q |V^{(l)}_{k-1}|$, which can be solved to yield
\begin{equation}
  \label{Q2}
 |V^{(l)}_k | = \frac{1}{2} q^{k-l}(q-1), \qquad l = 1 , \dots , k-1, \qquad |V^{(k)}_k| = q.
\end{equation}

The reason to consider the sets $V^{(l)}_k$ is that all the
elements of each such $V^{(l)}_k$ have the same degree
distribution, independent of $k$ for $l\leqslant k-1$. In fact,
the degrees of $i\in V^{(1)}_k$ are non-random since these nodes
receive no decorating bonds. For such $i$, $n(i) =
\sum_{j}n^{(0)}(j)$, where $n^{(0)}(j)$ is the degree of the
corresponding node in the basic pattern, and the sum is taken over
all such patterns which are glued up. By the symmetry of $M_1$\,,
$M_3$\,, and $M_5$\,, we have  $n(i) = 4$ for $M_1$ and
$M_3$\,, and $n(i) = 6$ for $M_5$\,. For $M_2$\,, $n(i)$ takes
values $3$, $4$, $5$, see figure~\ref{motif2}. For $i \in
V^{(l)}_k$, $l= 2, 3, \dots , k-1$, we have $n(i)= \tilde{n}(i) +
\nu(i)$, where $\tilde{n}(i)$ is non-random and has to be
calculated as just described. The summand $\nu(i)$ is the number
of decorating bonds attached to~$i$. To simplify our
consideration, let us stick to the case of the graph generated by
$M_1$\,. Then  for $l = 1 , \dots , k-1$ and $i\in V_{k}^{(l)}$,
we have $\tilde{n}(i) = 4$ and $\nu (i)$ takes values $ \nu = 0, 1
, 2 , \dots , 4 (l-1)$, with probability
\begin{equation}
  \label{Q3}
{\rm Prob} \left(\nu (i \right) = \nu)  = { 4(l-1) \choose \nu} p^\nu (1-p)^{4(l-1)-\nu}.
\end{equation}
For $i \in V^{(k)}_k$, $\nu (i)$ takes values $0, 1 , \dots ,
2(k-1)$. Therefore, the maximum node degree which can occur in
$V_k$\,, $k\geqslant 2$, is
\begin{equation}
  \label{Q3a}
 \max_{i \in V_k} n(i) = 4 k - 4.
\end{equation}
As is usual in the theory of real world networks, which are in
fact non-random, the randomness manifests itself as the random
choice of a node. If we apply this principle here, then~(\ref{Q3})
can be considered as the conditional probability distribution,
conditioned by the event that the node $i$ has been picked from
the set $V^{(l)}_k$. The probability  of the latter event is taken
to be proportional to the number of its elements, that is,
\begin{eqnarray}
  \label{Q4}
  {\rm Prob} \left( i \in V^{(l)}_k \right) & = & \frac{|V^{(l)}_k|}{|V_k|}
  = \frac{q-1}{1 + q^{1-k}} q^{-l} = \frac{2}{1 + 3^{1-k}} 3^{-l},
  \qquad l \leqslant k-1, \\[.2cm]
{\rm Prob} \left( i \in V^{(k)}_k \right) & = & \frac{2q}{q^k + q}
= \frac{2}{3^{k-1} + 1}\,. \nonumber
\end{eqnarray}
If we now take the expectation of $n(i)$ with respect to this distribution, that is,
\begin{eqnarray}
  \label{Q5}
 \langle n_k \rangle & = &  \sum_{l=1}^{k-1} \sum_{\nu =0}^{4(l-1)} ( 4 + \nu)
  \frac{2 \cdot 3^{-l}}{ 1 + 3^{1-k}}\cdot { 4(l-1)
  \choose \nu} p^\nu (1-p)^{4(l-1)-\nu}\\[.2cm] &
  + & \sum_{\nu=0}^{2 (k-1)} (2+ \nu) \frac{2}{3^{k-1} + 1}
  \cdot { 2(k-1) \choose \nu} p^\nu (1-p)^{2(k-1)-\nu}, \nonumber
\end{eqnarray}
we readily obtain $\langle n_k \rangle = 4 + 2 p -
{4(1+p)}/{(3^{k-1} + 1)} $, which is in full agreement with the
one given in~(\ref{n3}) or in table~\ref{characteristics}. In
order to figure out the limit $k\rightarrow +\infty$ of the
distribution given by~(\ref{Q3}) and~(\ref{Q4}) we calculate its
characteristic function, c.f.~(\ref{Q5}),
\begin{eqnarray}
  \label{Q6}
  \varphi_k (t)  & = &  \sum_{l=1}^{k-1} \sum_{\nu =0}^{4(l-1)}
  \exp\left[ {\rm i}t( 4 + \nu) \right] \frac{2 \cdot 3^{-l}}{ 1 +
  3^{1-k}}\,
  \cdot { 4(l-1) \choose \nu} p^\nu (1-p)^{4(l-1)-\nu}\\[.2cm]
  & + & \sum_{\nu=0}^{2 (k-1)}\exp\left[ {\rm i}t (2+ \nu) \right] \frac{2}{3^{k-1} + 1}\,
  \cdot { 2(k-1) \choose \nu} p^\nu (1-p)^{2(l-1)-\nu} \nonumber \\[.2cm]
  & = & \frac{2 e^{ 4 {\rm i} t}}{1 + 3^{1-k}}\, \cdot \frac{1 - 3^{1-k}
  \left(e^{{\rm i}t} p + 1 - p \right)^4 }
  { 3 -\left(e^{{\rm i}t} p + 1 - p \right)^4 }\nonumber \\[.2cm]
  & + & \frac{2 e^{2 {\rm i}t}}{3^{k-1} +1} \left(e^{{\rm i}t} p + 1 - p \right)^2 ,
  \qquad {\rm i} = \sqrt{-1}\,. \nonumber
\end{eqnarray}
Then the limiting characteristic function is
\begin{equation}
  \label{Q7}
\varphi (t) = \frac{2  e^{ 4 {\rm i} t}}{  3 -\left(e^{{\rm i}t} p
+ 1 - p \right)^4 }\,,
\end{equation}
which can be continued to a meromorphic function analytic in some
complex neighborhood of the point $t=0$. This means that the
limiting node degree distribution has all moments and hence cannot
be of scale-free type\footnote{For scale-free graphs, the node
degree distribution is  $P(k) = C k^{-\gamma}$, $k \geqslant 1$,
$\gamma>1$; hence,  $\sum_{k=1}^\infty k^m P(k)$  diverges for all
$m\geqslant \gamma -1$. }. This also agrees  with the heuristic
rule, see equation~(3.13) on page~188 in~\cite{Newman}, that for
scale-free networks
\begin{equation}
  \label{Q8}
  \max_{i \in V_k} n(i) \sim |V_k|^{1/(\alpha -1)}, \quad \alpha >1,
\end{equation}
whereas in our case we have~(\ref{Q3a}) and $|V_k | = (3^k +
3)/2$. In a similar way, one can show that all our graphs are not
scale-free. Another observation on this item can be made by
comparing the function~(\ref{Q7}) with the characteristic function
of the Poisson distribution~(\ref{j}) which has the form
\[\varphi_{\rm Poisson} (t) = \exp\left[c \left( e^{{\rm i}t} - 1 \right) \right],\]
and hence can be continued to a function analytic on the whole
complex plane. Therefore, the  distribution corresponding
to~(\ref{Q7}) with $p>0$ is intermediate as compared to the
Poisson and scale-free distributions. For $p=0$, the
function~(\ref{Q7}) is also entire.

\subsection{Amenability, clustering, and small world properties}

The next property of our graphs which we are going to address is
{\it amenability}. To introduce it we need one more notion. Let $G
= (V, E)$ be a countable graph with node set $V$ and bond set $E$.
For a finite $\mathit{\Delta} \subset V$, by $\partial
\mathit{\Delta}$ we denote the set of  nodes which are not in
$\mathit{\Delta}$ but have neighbors in $\mathit{\Delta}$. This
set is the {\it outer boundary} of $\mathit{\Delta}$, whereas the
elements of $\mathit{\Delta}$ which are neighbors to the elements
of $\partial \mathit{\Delta}$ constitute the {\it inner boundary}
of $\mathit{\Delta}$. As usual, by $|\mathit{\Delta}|$ and
$|\partial \mathit{\Delta}|$ we denote the number of nodes in
these sets. The graph $G$ is said to be {\it amenable} if there
exists a sequence of finite node sets $\{\mathit{\Delta}_k\}_{k\in
\mathbb{N}}$\,, such that
\begin{equation}
 \label{Q9}
\lim_{k\rightarrow +\infty}  \frac{|\partial \mathit{\Delta}_k|}{|\mathit{\Delta}_k|} = 0.
\end{equation}
If such a limit is positive for any sequence
$\{\mathit{\Delta}_k\}_{k\in \mathbb{N}}$\,, the graph is called
{\it nonamenabile}. The lattices $\mathbb{Z}^d$, $d\geqslant 1$,
are amenable graphs and for such sets one can take the cubes
\[
\mathit{\Delta}_k = \{ i = (i_1 , \dots i_d) \in \mathbb{Z}^d \ :
\ |i_j| \leqslant N_k\,, \ j= 1, \dots , d\},
\]
such that $N_{k+1} > N_k$ and $N_k \rightarrow + \infty$. In this
case, $|\mathit{\Delta}_k| \sim N_k^d$ and $|\partial
\mathit{\Delta}_k| \sim N_k^{d-1}$ and hence~(\ref{Q9}) holds.
Sometimes, sequences for which~(\ref{Q9}) holds are called {\it
Van Howe} sequences, see e.g.~\cite{Ruelle}. Cayley trees, except
for $\mathbb{Z}$, are nonamenable. Let us turn now to our graphs.
Due to their hierarchical structure, it is convenient to
check~(\ref{Q9}) for the sequence of node sets of $\Lambda_k$\,,
that is for $\{V_k\}_{k\in \mathbb{N}}$\,. By the construction of
$\Lambda_k$\,, the inner boundary of each $V_k$ is exactly the set
of all its external nodes, the number of which is equal to the
number of nodes in the corresponding motif, i.e. it is $q$. By
construction, one of them becomes an external node of
$\Lambda_{k+1}$\,, and the remaining $q-1$ ones become inner nodes
of $\Lambda_{k+1}$\,. For all the motifs $M_j$\,, $j=1, \dots ,
5$, we have the degree of any node therein being at most $3$,
see figure~\ref{motifs}. Then for all our graphs,
\[
\max_{i\in V_k} n(i) \leqslant 6 k ,
\]
c.f.~(\ref{Q3a}). At the same time, $|V_k| \sim q^k$, $q=3,4$,
see~(\ref{m2}), which immediately yields that all our graphs are
amenable.

Now we study the clustering in our graphs. For non-random graphs,
the clustering coefficient is defined as follows. For a given node
$i\in V$ of degree $n(i)$, let $N(i)$ be the number of bonds
linking its neighbors, which is the number of triangles with
vertex $i$. Clearly, $N(i) \leqslant n(i) [n(i)-1]/2$ and the
maximum value of this parameter is attained for complete graphs
where each node is a  neighbor to any other one. Thus, the
quantity
\begin{equation}
 \label{Q10}
Q(i) : = \frac{2 N(i)}{n (i) [ n(i) -1]}\,,
\end{equation}
characterizes clustering at node $i$. Then we define the clustering of our graphs  as
\begin{equation}
 \label{Q11}
Q = \lim_{k\rightarrow +\infty} \frac{1}{|V_k|} \sum_{i \in V_k} Q(i).
\end{equation}
Note that for many graphs, e.g., for trees or bipartite graphs,
one has $Q(i) =0$ for any node $i$, see also~\cite{EE,FU}. For
random graphs, the degree $n(i)$, as well as the parameter $N(i)$,
are random. Since in this case the calculation of $Q$ is much more
involved, we postpone it to the future. Here we only compare the
values of $Q$ obtained for the bare graphs with those for fully
decorated ones.

For the graph based on $M_1$ and a node $i\in V_k^{(l)}$, $l=1 ,
\dots , k-1$, we have: (a) $n(i) = 4$, $N(i) = 3$ for $l=1$, and
$N(i)=2$ for $l\geqslant 2$; (b) $ n(i)= 4l $, $N(i) = 4l$. Here (a) and (b)
correspond to a bare graph and to a fully decorated graph,
respectively. These numbers follow directly from the construction
of the graphs. The number of elements in each $V_k^{(l)}$ is given
in~(\ref{Q2}), which allows one to compute
\begin{eqnarray}
 \label{Q12}
{\rm (a)} \ & Q &= \lim_{k\rightarrow +\infty} \left(\frac{1}{3}
+ \frac{|V^{(1)}_k|}{6|V_k|} + \frac{2}{|V_k|} \right)
= \frac{4}{9} = 0.4444\ldots, \\[.2cm] \label{Q12a}
{\rm (b)} \ & Q & = 2\cdot3^{-1/4} \arctan{3^{-1/4}}-
\frac{1}{3^{1/4}} \ln \frac{3^{1/4} +1}{3^{1/4} -1}
\approx 0.525897.
\end{eqnarray}
For the bare graph based on $M_3$\,, we have $N(i) = 0$ for
all nodes; hence, $Q=0$ in this case. For the fully decorated
graph based on $M_3$\,, we observe that two nodes of $V^{(1)}_2$
have neighbors only in $V^{(1)}_2$, and the remaining four nodes
have two neighbors in $V^{(1)}_2$ and two~--- in $V^{(2)}_2$.
Since for the nodes $i\in V^{(1)}_2$, we have $N(i) =0$, like in
the case of the bare graph, we should consider these two groups
separately. Thus, we split each $V^{(l)}_k$, $l=1, \dots , k-1$,
into $V^{(l)}_{k,0}$ and $V^{(l)}_{k,1}$\,, where the first set
consists of the nodes which have neighbors only in
$V^{(l)}_{k,0}$\,. The elements of $V^{(l)}_{k,1}$ have also neighbors
in $V^{(l')}_{k}$ with $l' > l$. The number of nodes in these
sets are:
\[
 |V^{(l)}_{k,0}| = \frac{1}{2} 4^{k-l}, \qquad |V^{(l)}_{k,1}| =  4^{k-l}.
\]
For all $i\in  V^{(l)}_{k}$, $l=1, \dots , k-1$, we have $n(i) =
4l$. At the same time, $N(i) = 4 (l-1)$ for $i \in
V^{(l)}_{k,0}$\,, and $N(i) = 1 + 4 (l-1)$ for $i \in
V^{(l)}_{k,1}$\,. Putting all these numbers together we obtain
\begin{equation}
 \label{Q13}
Q = \frac{3}{2} \sum_{l=2}^\infty \frac{(l-1) 4^{-(l-1)}}{l (4l-1)} + \sum_{l=1}^\infty \frac{ 4^{- l}}{l (4l-1)} \approx 0.1223.
\end{equation}
For the bare graph based on $M_5$ one can obtain for $i \in
V_k^{(l)}, l=1,2,\dots, k-1$: $n(i)=6$,  $N(i)=8$ for $l=1$, and
$n(i)=6$, $N(i)=6$ for $l \geqslant 2$. For the fully decorated
graph based on $M_5$ we have for all internal nodes $n(i)=6l$, and
$N(i)=9$ for $l=1$, and $N(i)=12l+7$ for $l \geqslant 2$. Hence, we
obtain
\begin{eqnarray}
 \label{Q14}
 {\rm (a)} \ & & Q = \lim_{k\rightarrow +\infty} \left(\frac{2}{5}
 + \frac{2|V^{(1)}_k|}{15|V_k|} + \frac{12}{5|V_k|} \right) = 0.5, \\[.2cm]
{\rm (b)} \ & & Q  \approx 0.554145.  \nonumber
 \nonumber
\end{eqnarray}
There exists one more property of real world networks, which
Erd\H{o}s-R\'enyi type graph does not share, see
e.g.~\cite{Newman,BarratWeigt}. It is the so-called {\it
small-world property}, illustrated by Pal Erd\H{o}s himself in the
following way. All authors of mathematical papers are given the
{\it Erd\H{o}s} index which for his coauthors is equal to one. For
their coauthors who are not Erd\H{o}s' coauthors, this index is
equal to 2, and so on. Everyone whose papers are indexed by the
Mathematical Reviews can check his own index at the web-site
\href{http://www.ams.org/mathscinet/}{http://www.ams.org/mathscinet/}. It turns out that for many authors
it ranges from 2 to 6 (for the second named author of this paper
it is 4). To formulate the small-world property one needs the
following notion. A path in the graph is a sequence of nodes such
that every two consecutive elements of this sequence are neighbors
to each other. The length of the path is the number of such
consecutive pairs, which is equal to the number of bonds one
passes on the way from the origin to the terminus. If every two
nodes can be connected by a path, the graph is said to be
connected. For the given two nodes, $i$ and $j$, the length of the
shortest path which connects them is said to be the distance
$\rho(i,j)$ between these nodes. Informally speaking, a graph $G =
(V,E)$ has the small-world property (or is a small-world graph) if
every two nodes $i,j\in V$ are {``not too far''} from each other.
More precisely this property is formulated as follows. An infinite
graph $G$ has a small-world property if there exists a sequence of
its connected finite subgraphs $\{G_k\}_{k\in \mathbb{N}}$ with
the following property. Let ${\rm diam}(G_k) = \max_{i,j \in V_k}
\rho (i,j)$ be the {\it diameter} of $G_k$\,, $k\in \mathbb{N}$,
and $\langle n_k \rangle$ be the average value of the node degree
in $G_k$\,, that is, $\langle n_k \rangle = 2 |E_k| / |V_k|$. Then
the sequence $\{G_k\}_{k\in \mathbb{N}}$\,, and hence the graph
$G$, are said to have the small-world property if there exists a
positive constant $C$ such that for all $k\in \mathbb{N}$,
\begin{equation}
 \label{n4}
{\rm diam} (G_k) \leqslant C \log_{\langle n_k \rangle} |V_k|.
\end{equation}
This means that in such graphs, the distances between the nodes
scale at most logarithmically with the size of the graph. For our
graphs, the results on this item are given in the last three rows
of table~\ref{characteristics}. The diameters of $\Lambda_k$
presented therein were calculated in a routine way for the cases
where the graphs are bare ($p=0$) or are fully decorated ($p=1$).
In the former case, neither of our graphs has the small-world
property. However, this property holds for all fully decorated
graphs.

\begin{table}[ht]
\caption{The structural characteristics of the families of hierarchical
graphs.} \label{characteristics} 
\begin{center}
\renewcommand{\arraystretch}{2}
\begin{tabular}{|c|c|c|c|c|c|}
\hline
 motif   &$M_1$ & $M_2$ & $M_3$ & $M_4$ & $M_5$\strut\\
\hline $|V_k|$ & $\frac{3}{2}(3^{k-1}+1)$ &
        $2(4^{k-1}+1)$ & $2(4^{k-1}+1)$ &
        $2(4^{k-1}+1)$ & $2(4^{k-1}+1)$\strut\\
\hline
$|E_k|$ & $3^k$ & $4^k$& $4^k$ & $5\cdot 4^{k-1}$&$6\cdot 4^{k-1}$\strut\\
  & $\frac{3}{2}(3^{k-1}-1)p$ &
        $\frac{4}{3}(4^{k-1}-1)p$& $\frac{4}{3}(4^{k-1}-1)p$&
        $\frac{5}{3}(4^{k-1}-1)p$&$2(4^{k-1}-1)p$\strut\\
\hline

$\langle n_k \rangle$ & $4+2p$& $4+\frac{4}{3}p$ & $4+\frac{4}{3}p$ &
    $5+\frac{5}{3}p$ & $6+2p$  \strut\\

 & $-4\frac{1+p}{3^{k-1}+1}$ &
        $-\frac{4}{3}\frac{3+2p}{4^{k-1}+1}$ & $-\frac{4}{3}\frac{3+2p}{4^{k-1}+1}$ &
        $-\frac{5}{3}\frac{3+2p}{4^{k-1}+1}$ & $-2\frac{3+2p}{4^{k-1}+1}$ \strut\\
\hline
$\mathrm{diam}(\Lambda_k)$,  $p=0$ & $2^{k-1}$ & $2^k$ & $2^k$ & $2^k$ & $2^{k-1}$ \strut\\
\hline
$\mathrm{diam}(\Lambda_k)$,  $p=1$ & $k$ & $k+1$ & $2(k-1)$ & $k$ & $k$ \strut\\
\hline
$C$, $p=1$ & $(\log_6 3)^{-1}$ & $(\log_6 4)^{-1}$ & $2(\log_6 4)^{-1}$ &
        $(\log_7 4)^{-1}$ & $(\log_8 4)^{-1}$\strut\\
\hline
\end{tabular}
\renewcommand{\arraystretch}{1}
\end{center}
\end{table}
In table~\ref{characteristics}, we  collected the structural
characteristics of our graphs based on the motifs
$M_1$\,,~\ldots~, $M_5$  given in figure~\ref{motifs}. Its first
three rows are obtained from the formulas~(\ref{m2})
and~(\ref{n3}).

\section{Graph structure and the Ising model properties}

As was mentioned above, there exists a profound connection between the
properties of Gibbs random fields of models based on graphs and
the structural properties of these graphs\footnote{Of course, we
are speaking now about countably infinite graphs.}. For the
hierarchical lattices, the notion of the Gibbs random field of the
Ising model was introduced in~\cite{BZ,BleherZalys1989}. In the
physical terminology, each (pure) Gibbs random field corresponds
to a state of thermal equilibrium of the model, see~\cite{G} for
more details. Accordingly, the existence of multiple Gibbs random
fields corresponds to the existence of multiple equilibrium states
and hence of phase transitions. If there is no interaction between the
spins, the Gibbs random field is unique. However, if the
interaction is strong enough and if it is  effectively propagated
by the underlying graph (high enough ``connectivity''), the Gibbs
fields can be multiple. The  condition of high connectivity is
essential, which can be seen from the fact that the Ising model on
the one-dimensional lattice $\mathbb{Z}$, considered as a graph
with the natural adjacency relation, bears only one Gibbs field
and hence no phase transitions can occur in this case -- no matter
how strong the interaction is. With regard to these arguments, we
can divide all graphs into three groups according to the following
property of the Ising model defined on these graphs: \vskip.1cm
\begin{tabular}{ll}
(a) \ & There exists only one Gibbs random field for all interactions.\\[.2cm]
(b) \ & There exists only one Gibbs random field for weak interactions, and multiple Gibbs\\[.1cm] & random fields for strong interactions.\\[.2cm]
(c) \ & There exist multiple Gibbs random fields for all nonzero interactions.
\end{tabular}
\vskip.1cm  \noindent As was mentioned above, the lattice
$\mathbb{Z}$ belongs to the first group. The lattices
$\mathbb{Z}^d$ with $d\geqslant 2$, as well as all Cayley
trees\footnote{Except for $\mathbb{Z}$, see page~247 in~\cite{G}.}
belong to the second group. An example of the graph belonging to
the third group can be found in section~4 of~\cite{KK}. Clearly,
the above classification should be made more precise for random
graphs, since in this case the existence of a Gibbs random field is a
random event. We return to this issue below.

The Ising model defined on the graphs from the second group may
have a {\it critical point}, which separates the regime of
uniqueness (weak interactions) from the regime of non-uniqueness
(strong interactions). At their critical point, the Gibbs random
fields have ``unusual'' properties, which can be detected without
explicit construction of these fields. One of such properties is
the so-called {\it self-similarity}, which manifests itself in the
appearance of unstable fixed points of some (renormalization)
transformations. It turns out that the hierarchical graphs are
quite suitable for such a study~--- that it why they appear in
this context. Thus, one can have the following criterion: if the
Ising model has a critical point, the graph belongs to the second
group. If not,~then the graph belongs either to the first or to the
third group. The latter cases can be distinguished by an
additional study.

In contrast to the hierarchical lattices introduced and studied
in~\cite{HinczBerker2006}, our bare graphs belong to the first
group. This can be seen from the analysis which we present
below~--- a direct proof of such a statement will be done in our
forthcoming publication. Thus, the interaction along the
decorating bonds plays the main role in the possible appearance of
 phase transitions in the Ising model defined on our graphs. In
view of this fact, we will assume that the solid bonds bear the
interaction $K$, whereas the dotted bonds bear the interaction
$L$. As the  bonds of the latter kind  appear at  random, we can
think of the corresponding model as of the one defined on the
fully decorated graph with random interactions along the
decorating bonds. Thus, in this model, the interaction along each
 bond $\langle i , j\rangle$ is a random variable, denoted
by $L^\omega_{ij}$\,, which takes values  $0$ and some nonzero
$L\in \mathbb{R}$ with probabilities $1-p$ and $p$, respectively.
For different bonds, these random variables are independent.

For a given $k\in \mathbb{N}$, the Ising model on the fully decorated graph $\Lambda_k$ is defined in the usual
way by assigning spin variables $\sigma_i=\pm 1$ to the nodes $i \in
V_k$ and by setting the Hamiltonian
\begin{equation} \label{n7}
  -\beta\mathcal{H}_k = h \sum_{i\in V_k} \sigma_i
  + K \sum_{\langle i,j \rangle \in E_k'} \sigma_i\sigma_ j
  + \sum_{\langle i,j \rangle \in E_k''} L^\omega_{ij}  \sigma_i\sigma_ j\,, \quad
                      \quad h,K\in \mathbb{R}, \ \ k\in \mathbb{N},
\end{equation}
where $h$ is an external field. In accordance with the above
arguments, the third summand corresponds to the interaction along
decorating bonds. We have included the first term in view of the
following arguments. As is known, the Ising model on the lattices
$\mathbb{Z}^d$, $d\geqslant 2$, exhibits phase transitions only if
$h=0$. This is also the case if the graph is amenable and
quasi-transitive. For nonamenable graphs, this model may have a
phase transition for nonzero $h$, see~\cite{JS}.

By the construction of our graphs, each $\Lambda_k$ has $q$
external nodes, $i_1 , \dots, i_q$\,, and the corresponding number
of internal nodes, see table~\ref{characteristics}. Such external
nodes can be considered as a boundary of $\Lambda_k$\,.
Correspondingly, we call the spin variables {\it external} or {\it
boundary} (respectively, {\it internal}) spins if they are
assigned to external (respectively, internal) nodes. For a fixed
configuration of the external  spins $\sigma_{i_1}\, , \dots ,
\sigma_{i_q}$\,, we consider
\begin{equation}
  \label{N1}
Z^\omega_k (\sigma_{i_1}\, , \dots , \sigma_{i_q}) = \sum_{{\rm
internal} \ {\rm spins } \ {\rm of} \ \Lambda_k} \exp\left(
-\beta\mathcal{H}_k \right).
\end{equation}
This is the conditional partition function of the spin system in
$\Lambda_k$\,, conditioned by the fixed configuration of spins on
the boundary of $\Lambda_k$\,. Of course, it depends on the graph,
i.e. on the motif used in its construction. According to the main
principle of the theory of Gibbs random fields applied in our
context, see~\cite{G}, the fact that such a field is unique is
ensured by the conditional partition function asymptotic
independence, as $k\rightarrow +\infty$, of the boundary spins.
For our graphs, $Z^\omega_k$ are random and hence also depend on
$\omega$. In this situation, one can apply the following two
approaches: the above mentioned independence holds (i) for almost
all $\omega$; (ii) in average. The latter corresponds to the fact
that the expected values of $Z^\omega_k$\,, which we denote by
$\langle Z^\omega_k \rangle$, are independent of the boundary
spins. These two approaches are called {\it quenched} and {\it
annealed} disorders, respectively, see~\cite{Bovier}, page~99. We
will work in the annealed approach. It can be shown that the
summands in~(\ref{N1}) are positive and separated from zero. Since
the number of these summands  increases to $+\infty$ as
$k\rightarrow +\infty$,  we have  $\langle Z^\omega_k \rangle
\rightarrow +\infty$. Therefore, a weaker form of the uniqueness
condition can be
\begin{equation}
  \label{N2}
 \lim_{k\rightarrow +\infty} \langle Z^\omega_k (\sigma_{i_1}\,, \dots ,
 \sigma_{i_q}) \rangle/\langle Z^\omega_k (\sigma'_{i_1}\,, \dots , \sigma'_{i_q}) \rangle = c,
\end{equation}
which holds for any two configurations of the boundary spins and
for some $c>0$, which depends on these configurations. The
convergence in~(\ref{N2}) should be stable with respect to small
changes of $h$ and $K$, i.e. of the starting element $Z_1$\,. In
this case, the annealed limiting  free energy per spin is
independent of the boundary configuration. Correspondingly, a
necessary condition for the latter quantity to be dependent on the
boundary spins is that
\begin{equation}
  \label{N3}
 \lim_{k\rightarrow +\infty} \langle Z^\omega_k (\sigma_{i_1}\,, \dots , \sigma_{i_q})
 \rangle/\langle Z^\omega_k (\sigma'_{i_1}\,, \dots , \sigma'_{i_q}) \rangle = +\infty,
\end{equation}
for some $\sigma_{i_1}\,, \dots , \sigma_{i_q}$ and
$\sigma'_{i_1}\,, \dots , \sigma'_{i_q}$\,. We say that the Ising
model defined on our graph has a critical point, if  there exists
$c'>0$, distinct from $c$ in~(\ref{N2}), such that for  $K=K_*$
and $h=h_*$\,,
\begin{equation}
  \label{N4}
   \lim_{k\rightarrow +\infty} \langle Z^\omega_k
   (\sigma_{i_1}\,, \dots , \sigma_{i_q}) \rangle/\langle Z^\omega_k
   (\sigma'_{i_1}\,, \dots , \sigma'_{i_q}) \rangle = c'.
\end{equation}
Here $K_*$ and $h_*$ are certain values of the interaction
intensity and the external field. The convergence in~(\ref{N4})
should be such that for arbitrarily small deviations from the
point $(K_*\,,h_*)$ one has either~(\ref{N2}) or~(\ref{N3})
instead of~(\ref{N4}). Note that we do not exclude that $h_*\neq
0$ since our graphs are not quasi-transitive and hence the result
of~\cite{JS} may not be applicable in this case.

\section{Critical points of the Ising model on $M_1$ based graph}

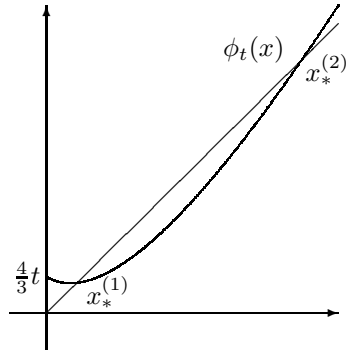
\begin{figure}
\unitlength 0,24mm
\centering
\begin{picture}(150, 180)
    \put(0,20){\vector(1,0){180}}
    \put(20,0){\vector(0,1){190}}
    \put(20,20){\line(1,1){160}}
   \qbezier(20,40)(65,13)(180,190)
    \put(2,36){$\frac{4}{3}t$}
    \put(42,25){$x_*^{(1)}$}
    \put(162,148){$x_*^{(2)}$}
    \put(117,162){$\phi_t(x)$}
\end{picture}
\qquad\qquad\qquad\qquad
 \caption{Graphical solution of~(\ref{equS}).}
\label{wykres}
\end{figure}

For $k=1$, we have only basic bonds and no inner nodes. Thus,
\begin{equation}
  \label{CC1}
Z_1(a, b, c) = \exp \left[ - \beta \mathcal{H}_1 (a, b, c) \right]
= \exp[K(ab + ac + bc )+ h(a+b+c)],
\end{equation}
where we use the same notations for the nodes $a, b, c$, see the
left-hand graph in figure~\ref{construction}, as well as for the
corresponding spin variable, that is, for short we write $a =
\sigma_a$\,, $b = \sigma_b$\,, ect. For $k=2$, we have, see the
middle graph in figure~\ref{construction},
\[
- \beta \mathcal{H}_2 (a,b,c, \alpha, \beta , \gamma) =    - \beta \mathcal{H}_1 (a, \gamma, \beta)  - \beta \mathcal{H}_1 (\gamma, b, \alpha)  - \beta \mathcal{H}_1 (\beta, \alpha, c) + L^\omega_{ab} a b + L^\omega_{bc}  bc + L^\omega_{ac} a c.
\]
Thus, to obtain $Z_2(a, b, c)$ we have to sum out the internal
spins, see~(\ref{N1}). That is,
\[
Z^\omega_2(a, b, c) = \exp\left[L_{ab}^\omega ab + L_{ac}^\omega ac
                                       + L_{bc}^\omega bc\right] \sum_{\alpha , \beta , \gamma} Z_1(a,  \gamma, \beta) Z_1 (\gamma, b, \alpha) Z_1 (\beta, \alpha, c),
\]
where the sum is taken over $\alpha, \beta, \gamma = \pm 1$.
According to the hierarchical structure of the underlying graph,
for all $k\geqslant 2$ the graph $\Lambda_k$ is obtained from the
subgraphs $\Lambda_{k-1}$ exactly in the same way. This yields the
following recurrence for the partition
functions~(\ref{N1})
\begin{equation} \label{n9}
Z^\omega_k(a, b, c) = R_k^\omega(a, b, c) \sum_{\alpha , \beta , \gamma} Z_{k-1}(a,  \gamma, \beta) Z_{k-1} (\gamma, b, \alpha) Z_{k-1} (\beta, \alpha, c),
\end{equation}
where
\begin{equation} \label{n10}
R_k^\omega(a, b, c) =\exp\left[L_{ab}^\omega ab + L_{ac}^\omega ac
                                       + L_{bc}^\omega bc\right],
\end{equation}
and $Z_1 (a,b,c)$ is given in~(\ref{CC1}). From the independence
of $L^\omega_{ij}$ for different bonds, it follows that the random
variables $Z_{k-1}^\omega$ on the right-hand side of~(\ref{n9})
are independent, which allows us to compute the expectations and
obtain
\begin{equation} \label{n13}
  \langle Z_k^\omega(a, b, c)\rangle = \langle R_k^\omega(a, b, c) \rangle
                        \sum_{\alpha, \beta, \gamma}
                        \langle Z_{k-1}^\omega(a, \beta, \gamma) \rangle
                        \langle Z_{k-1}^\omega(\alpha, b, \gamma) \rangle
                        \langle Z_{k-1}^\omega(\alpha, \beta, c) \rangle.
\end{equation}
We recall that the random variables $L^\omega_{ij}$ take the
values $L\neq 0$ and $0$ with probabilities $p$ and $1-p$,
respectively. In view of the  model symmetry, c.f.~(\ref{CC1}),
the system~(\ref{n13}) involves the following variables
\begin{eqnarray}\label{xy3}
    A_k &=& \langle Z^\omega_k(1, 1, 1) \rangle,
    \quad \quad \,\,\,\, B_k = \langle Z^\omega_k(1, 1, -1)\rangle,\\[.2cm]
    C_k &=& \langle Z^\omega_k(-1,- 1, 1) \rangle,
    \quad D_k = \langle Z^\omega_k(-1,-1, -1)\rangle, \nonumber
\end{eqnarray}
which have the initial values
\begin{equation}
  \label{XY4}
 A_1 = e^{3(K+h)}, \ \quad B_1 = e^{-K + h}, \ \quad C_1 =  e^{-K - h}, \ \quad D_1 =  e^{3(K - h)}.
\end{equation}
By direct calculations,
\begin{eqnarray}
  \label{XY5}
  \langle R_k^\omega(1, 1, 1) \rangle & = & (p\,\exp(L)+ 1-p)^3, \\[.2cm]
 \langle R_k^\omega(-1, 1, 1) \rangle & = & (p\,\exp(L)+ 1-p)(p\,\exp(-L)+ 1-p)^2. \nonumber
\end{eqnarray}
Thereafter, the recursion~(\ref{n13}) can be written in the form
\begin{eqnarray}\label{uklad3}
x_{k+1} & = & t \frac{P_x (x_k , y_k )}{Q(x_k , y_k , z_k)}\,, \\[.2cm]
y_{k+1} & = &  \frac{P_y (x_k , y_k, z_k )}{Q(x_k , y_k , z_k)}\,, \nonumber \\[.2cm]
z_{k+1} & = & t \frac{P_z ( y_k, z_k )}{Q(x_k , y_k , z_k)}\,,
\nonumber
\end{eqnarray}
with the initial conditions
\begin{equation}
  \label{XY6}
x_1 = e^{4(K+h)}, \ \qquad y_1 = e^{2h} , \ \qquad z_1 =  e^{4K -
2 h},
\end{equation}
where we have used the notations
\begin{equation}
  \label{XY7}
x_k = A_k/C_k\,,  \qquad y_k = B_k/C_k\,,  \qquad z_k = D_k/C_k\,,
 \qquad k \in \mathbb{N},
\end{equation}
and
\begin{equation}
  \label{XY8}
 t = \left(\frac{p \,e^L + 1-p}{p\, e^{-L} + 1-p}\right)^2.
\end{equation}
The polynomials in~(\ref{uklad3}) have the following form
\begin{eqnarray}
  \label{XY9}
 P_x (x, y) & = & x^3 + 3 x y^2 + 3 y^2 +1, \\[.2cm]
P_y (x, y,z) & = & x^2 y + y^3 + 2 xy + y^2 z + 2 y + z, \nonumber \\[.2cm]
P_z ( y,z) & = & y^3 + 3 y + 3 z + z^3, \nonumber \\[.2cm]
Q(x,y,z) & = & x y^2 + x + 2 y^2 + 2 yz + z^ 2 + 1 . \nonumber
\end{eqnarray}
For every $t>0$, the mapping $(x_k\, , y_k\, , z_k) \mapsto
(x_{k+1}\, , y_{k +1}\,, z_{k+1}) = T_t(x_k\, , y_k\, , z_k)$\,,
defined in~(\ref{uklad3}) and~(\ref{XY9}), maps  the first octant
of $\mathbb{R}^3$ into itself. The starting point $(x_1\,, y_1\,,
z_1)$ lies on the surface $\mathcal{S}$ defined by the equation,
c.f.~(\ref{XY6}),
\begin{equation}
  \label{XY10}
 x = y^3 z,
\end{equation}
which, however, is not preserved by this mapping, which can be
checked directly. Let $\mathcal{T}_t$ be the invariant set of the
map $T_t$\,, that is, the set of points $(x,y,z)$ such that $T_t
(x, y , z) = (x, y, z)$. Suppose that the intersection of
$\mathcal{T}_t$ with $\mathcal{S}$ is non-void. If $(x_*\,, y_*\,,
z_*)$ belongs to this intersection and $(x_1\, , y_1\,, z_1)
=(x_*\,, y_*\,, z_*)$\,,  then $(x_k\, , y_k\,, z_k) =(x_*\,,
y_*\,, z_*)$ for all $k\in \mathbb{N}$. Let us find such points.
From the second equation in~(\ref{uklad3}) for $x = y^3 z$ we get
\[ y Q( y^3 z, y , z) = P_y( y^3 z, y, z),
\]
which can be transformed into the following
\begin{equation}
  \label{XY11}
y^3  (1+ y^3 z) ( 1 - yz) = (y+z) (1-yz).
\end{equation}
A solution of the latter equation is $z = 1/y$, thus $x = y^2$,
which being used in the first equation in~(\ref{uklad3}) yields
$(1+ y^2)^2 = t (1+ y^2)^2$ and hence $t = 1$. For $p>0$, the
latter implies $L=0$,  and also $K=0$, see~(\ref{XY6}). This fixed
point corresponds to a noninteracting system of spins in an
external field and thereby has nothing to do with critical points.
For $z \neq 1/y$, from~(\ref{XY11}) we get
\begin{equation}
  \label{XY12}
 z (1-y^2) (y^4 + y^2 +1) = - y (1-y^2),
\end{equation}
which for positive $y$ and $z$ has a unique solution $y =1$, and
hence $ x = z$. We insert this into the first equation in
(\ref{uklad3}) and obtain
\begin{equation}\label{equS}
  x =  t\frac{x^3 + 3x + 4}{x^2 + 4x + 3} = t \frac{x^2 - x + 4}{x+3} := \phi_t(x).
\end{equation}
For $t =1$, the only solution is $x=1$, which corresponds to
$L=K=0$. The choice $t=1$ also corresponds to a bare graph
($p=0$); hence, the bare graph based on $M_1$ has no critical
points and no phase transitions. As we have already mentioned
above, in a separate work we prove that the Gibbs states of the
Ising model based on this graph are always unique. For $t<1$, that
is, for $L<0$,~(\ref{equS}) has only one positive solution
\begin{equation}
  \label{XY13}
 x_* = \frac{ - 3 - t + \sqrt{ 9 + 22 t - 15 t^2}}{ 2 (1-t)} < 1,
\end{equation}
which corresponds to $K_* < 0$. It is stable and $(x_k\,, y_k\, ,
z_k) \rightarrow (x_* , 1 , x_*)$ as $k\rightarrow +\infty$, for
any $(x_1\,, y_1\, , z_1) \in \mathcal{S}$. Thus, for $L<0$, the
Ising model has no critical points and no phase transitions, even
for the fully decorated graph. This can be caused by a
frustration, which takes place in an antiferromagnetic Ising model
on triangles. For $t>1$, figure~\ref{wykres} presents the
graphical solutions of~(\ref{equS}).
They have the following form
\begin{equation} \label{m18}
    x^{(1)}_* = \frac{3+t-\sqrt{9+22t-15t^2}}{2(t-1)}\,,
    \qquad x^{(2)}_* = \frac{3+t+\sqrt{9+22t-15t^2}}{2(t-1)}\,.
\end{equation}
By direct calculations, we get  $\phi_t'(x^{(1)}_*)< 1$ and
$\phi_t'(x^{(2)}_*)> 1$, see figure~\ref{wykres}. Hence,
$x^{(1)}_*$ is stable, whereas  $x^{(2)}_*$ is unstable. Note also
that $x^{(1)}_* \rightarrow 1$ and $x^{(2)}_* \rightarrow \infty$
as $t \rightarrow 1$. The solutions~(\ref{m18}) exist and are
considered to be distinct, provided $t \in (1, 9/5)$. This yields the condition
\begin{equation} \label{m19}
p < \psi (L) := \frac{3-\sqrt{5}}{\sqrt{5}\exp(L) - 3\exp(-L) +
3-\sqrt{5}}\,.
\end{equation}
Since $\psi (L)$ is a decreasing function,  the equation $\psi (L) = 1$ has only one solution
\begin{equation} \label{m20}
L_* = \frac{1}{4} \ln \frac{9}{5}\,.
\end{equation}
Then, for $L\leqslant L_*$\,, the critical point exists  for all $p
\in (0,1]$. For such $L$ and $p$, let $K_*(L,p)$ be the solution
of the equation
\begin{equation}
  \label{m21}
\exp(4 K) =   x^{(2)}_*.
\end{equation}
Then
\begin{equation}
  \label{m22}
  \lim_{k \rightarrow +\infty} x_k =\left\{ \begin{array}{ll} x^{(1)}_*
  \quad &{\rm if} \ \ x_1 < \exp[4K_* (L,p)], \\[.2cm] x^{(2)}_*
  \quad &{\rm if} \ \ x_1 = \exp[4K_* (L,p)], \\[.2cm] +\infty \quad &{\rm if} \ \ x_1 > \exp[4K_* (L,p)].\end{array} \right.
\end{equation}
The conditions in~(\ref{m22})  can be formulated directly for $K$,
e.g. $x_k \rightarrow x^{(1)}_*$ if $K < K_*(L,p)$. Thus,
$x^{(1)}_*$ is the so-called high-temperature fixed point. For $L
> L_*$\,, we have $\psi(L) < 1$. Hence, the condition~(\ref{m19}) is
not satisfied for $p\in (\psi (L), 1]$, which  also includes the
fully decorated graph with $p=1$. For such $L$ and $p$, the whole
graph of $\phi_t$ lies  above the line $\phi_t(x) =x$, which means
that $x_k \rightarrow +\infty$ for all initial $x_1\geqslant 0$.
Thus, for $L>L_*$ and $p\in (\psi(L_*),1]$, the model has no
critical points and the Gibbs random fields are multiple for all
$K\in \mathbb{R}$. For $p=\psi (L_*)$, we have $t = 9/5$ and
$x^{(1)}_* = x^{(2)}_* =3$, which corresponds to $K_* = (\ln
3)/4$. In this case, $x_k \rightarrow 3$ if $K \leqslant K_*$\,,
and $x_k \rightarrow +\infty$ if $K > K_*$\,.

\section{Concluding remarks}

\subsection{Construction and structural properties}

The graphs introduced in this paper are constructed from motifs
$M_1\,, \dots, M_5$ in an algorithmic way, like the hierarchical
lattices introduced in~\cite{GriffithsKaufman1982}, and employed
in~\cite{HinczBerker2006} where they were supplied with long-range
(decorating) bonds. However, our main construction  principle~---
to replace nodes with the graphs of the previous level~--- differs
from the one used in those papers, where the graphs of the
previous level replaced the bonds. Hierarchical graphs of this kind
model the real world networks with the so-called modular structure,
which are organized in tightly knit communities with relatively
sparse connections between them. For more details herein, we
refer the reader to~\cite{hincz,modul} and to the publications
cited therein. In contrast to the graphs introduced
in~\cite{HinczBerker2006}, see also~\cite{hincz}, our decorated
graphs are not scale-free. For them, the node degree distribution
is of intermediate type such that all the moments exist. The
corresponding characteristic function can be extended to a
meromorphic function, analytic in the complex neighborhood of
zero, see~(\ref{Q7}), whereas for scale-free graphs such functions
are nonanalytic at zero. All our graphs are amenable, which
correlates with the property just discussed,
see~(\ref{Q3a}),~(\ref{m2}), and~(\ref{Q8}). An interesting property is that, like for
the graphs studied in~\cite{HinczBerker2006}, adding decorations
forms the graphs possessing the small-world property.

\subsection{Critical points}

The appearance of critical points, and hence of phase transitions, for
the Ising model confirms good communicating properties of the
underlying graphs. For our graphs, in contrast to the hierarchical
lattices for which phase transitions take place even without
decorations, critical points can exist only if the decorations are
applied, with any $p>0$. The most interesting related fact is that
for the graph based on $M_1$\,, the antiferromagnetic Ising model
has no phase transitions for any $L$ and $p$. This is due to the
generic frustration of such models~--- the elementary factor
$R_k(a,b,c)$ which appears in~(\ref{n10}) with negative $L$ cannot
be maximized. We expect that the same is true for the graphs based
on $M_5$. To check if this conjecture of ours is true we plan to study
a Potts model on the same graph and $q\geqslant 3$, for which such
a frustration is no longer actual. We also plan to look for
critical points with nonzero $h$, as well as for critical points
of the  Ising model with  the underlying graphs based on the
remaining motifs. However, the corresponding problems are much
tougher, so that the appropriate numerical means should be
applied.

\section{Acknowledgement}

The authors benefited from
fruitful discussions on the matter of this work held with our
colleagues Yuri Kondratiev and Vasyl Ustimenko for which they are
cordially grateful. The authors are also grateful to the unnamed
referee whose remarks and suggestions were helpful in improving
the presentation of the article.

\ukrainianpart

\title{▓║ЁрЁї│ўэ│ тшярфъют│ уЁрЇш ч ьюЄшт│т: ёЄЁєъЄєЁэ│ тырёЄштюёЄ│ Єр ъЁшЄшўэ│ Єюўъш ьюфхы│ ▓ч│эур}
\author{╠. ╩юЄюЁютшў, ▐Ё│щ ╩ючшЎ№ъшщ}
\address{
▓эёЄшЄєЄ ьрЄхьрЄшъш, ╙э│тхЁёшЄхЄ ╠рЁ│┐ ╩■Ё│-╤ъыюфютё№ъю┐, ╦■сы│э}

\makeukrtitle

\begin{abstract}
\tolerance=3000%
┬тюфшЄ№ё  │ тштўр║Є№ё  ъырё тшярфъютшї уЁрЇ│т, чсєфютрэшї т рыуюЁшЄь│ўэшщ ёяюё│с ч я' Єш ьюЄшт│т, чэрщфхэшї є [Milo~R., Shen-Orr~S., Itzkovitz~S., Kashtan~N., Chklovskii~D., Alon~U., Science, 2002, {\bf 298}, 824]. ╩юэёЄЁєъЎ│щэр ёїхьр эрурфє║ ёїхьє, чрёЄюёютрэє є [Hinczewski~M., A.~Nihat Berker, Phys. Rev.~E, 2006, {\bf 73}, 066126], чу│фэю ч  ъю■ ъюЁюЄъюё цэ│ ЁхсЁр ║ эхтшярфъют│, Єюф│  ъ фютуюё цэ│ ЁхсЁр тшэшър■Є№ эхчрыхцэю │ ч юфэръютю■ щьют│Ёэ│ёЄ■. ╬яшёрэю Ё ф ёЄЁєъЄєЁэшї тырёЄштюёЄхщ уЁрЇ│т, ёхЁхф  ъшї ║ Ёючяюф│ы ёЄєяхэ│т, ъырёЄхЁэ│ёЄ№, рьхэрс│ы№э│ёЄ№, тырёЄшт│ёЄ№ Є│ёэюую ёт│Єє. ─ы  юфэюую ч ьюЄшт│т тштўр║Є№ё  ъЁшЄшўэр Єюўър ьюфхы│ ▓ч│эур, тшчэрўхэю┐ эр т│фяют│фэюьє уЁрЇ│.
\keywords  рьхэрс│ы№э│ёЄ№, Ёючяюф│ы ёЄєяхэ│т, ъырёЄхЁэ│ёЄ№,  уЁрЇ Є│ёэюую ёт│Єє, ьюфхы№ ▓ч│эур, ъЁшЄшўэр Єюўър
\end{abstract}


\begin{thebibliography}{99}
     \bibitem{ErdosRenyi1959} Erd\H{o}s~P., R\'{e}nyi~A.,
     Publicationes Mathematicae, 1959, \textbf{6}, 290.
     \bibitem{ErdosRenyi1960} Erd\H{o}s~P., R\'{e}nyi~A., Publications of the
Mathematical Institute of the Hungarian Academy of Sciences,
1960,\textbf{5}, 17.
\bibitem{Burda1} Bia{\l}as,~P., Burda,~Z., Wac{\l}aw,~B.
Causal and homogeneous networks. In: {Science of complex
networks,}  14,
AIP Conf. Proc., \textbf{776}, Amer. Inst.
Phys., Melville, NY, 2005.

     \bibitem{WattsStrogatz1998} Watts~D.J., Strogatz~H., Nature, 1998,
     \textbf{393}, 440; 
      \bibdoi{10.1038/30918}.
     \bibitem{BarabasiAlbert1999} Barab\'{a}si~A.-L., Albert~R., Science, 1999,
     \textbf{286}, 509; 
     \bibdoi{10.1126/science.286.5439.509}.

    \bibitem{Milo} Milo~R. et al., Science, 2002, \textbf{298},
    824; 
    \bibdoi{10.1126/science.298.5594.824}.
     \bibitem{ItzkovitzMiloKashtan2003} Itzkovitz~S. et al.,
     Phys. Rev.~E, 2003, \textbf{68}, 026127; 
     \bibdoi{10.1103/PhysRevE.68.026127}.
\bibitem{Alon} Alon~U., Science, 2003, \textbf{301}, 1866; 
\bibdoi{10.1126/science.1089072};
Nature Reviews Genetics, 2007, \textbf{8}, 450; 
\bibdoi{10.1038/nrg2102}.
    \bibitem{Kashtan} Kashtan~N., Itzkovitz~S., Milo~R., Alon~U., Phys. Rev.~E,
2004, \textbf{70}, 031909;\\ 
\bibdoi{10.1103/PhysRevE.70.031909}.
    \bibitem{ItzkovitzAlon} Itzkovitz~S., Alon~U., Phys. Rev.~E, 2005,
\textbf{71}, 026117; 
\bibdoi{10.1103/PhysRevE.71.026117}.

\bibitem{Matias} Matias~C. et al,
REVSTAT, 2006, \textbf{4}, 31. 


\bibitem{Hag} H\"aggstr\"om~O., {Adv. Appl. Probab.,} 2000, \textbf{32}, 39. 

    \bibitem{GriffithsKaufman1982} Griffiths~R.B., Kaufman~M., Phys. Rev.~B,
    1982, \textbf{26}, 5022; 
    \bibdoi{10.1103/PhysRevB.26.5022}.
   \bibitem{BZ} Bleher~P.M., \v{Z}alys~E., Litovsk. Mat. Sb., 1988,
   {\bf 28},  252 
   [English translation in  Lithuanian Math.~J., 1989, {\bf 28},
   127; 
   \bibdoi{10.1007/BF01027189}].
    \bibitem{BleherZalys1989} Bleher~P.M., \v{Z}alys~E.,
    Commun. Math. Phys., 1989, \textbf{120}, 409; 
    \bibdoi{10.1007/BF01225505}.
    \bibitem{HinczBerker2006} Hinczewski~M., A.~Nihat Berker, Phys. Rev.~E,
    2006,  \textbf{73}, 066126; 
    \bibdoi{10.1103/PhysRevE.73.066126}.

    \bibitem{Wrobel} Wr\'{o}bel~M., Condens. Matter Phys., 2008, \textbf{11}, 341. 
     \bibitem{Kotorowicz} Kotorowicz~M., Albanian Journal of Mathemathics,
     2008, \textbf{3}, 235. 

\bibitem{Barlow} Barlow~M.T. Heat kernels and sets with fractal structure.
In: { Heat kernels and analysis on manifolds, graphs, and metric
 spaces (Paris, 2002).} Contemp. Math., 338, Amer. Math. Soc., Providence, RI, 2003.

    \bibitem{Newman} Newman~M.E.J., SIAM Review, 2003, \textbf{45},
    167; 
    \bibdoi{10.1137/S003614450342480}.
 \bibitem{Ruelle} Ruelle~D., Statistical mechanics: Rigorous results.
 W.A.~Benjamin, Inc., New York--Amsterdam, 1969.
\bibitem{EE}  Erd\H{o}s~P., Kleitman~D.J.,  Rothschild~B.L., Asymptotic
enumeration of $K_n$-free graphs. In: {Colloquio Internazionale
sulle Teorie Combinatorie} (Rome, 1973), Tomo II, pp. 19. 
Atti dei Convegni Lincei, No. 17, Accad. Naz. Lincei, Rome, 1976.

\bibitem{FU} Futorny~V.,  Ustimenko~V.,  Acta Appl. Math., 2007, \textbf{ 98},
47; 
\bibdoi{10.1007/s10440-007-9144-8}.

    \bibitem{BarratWeigt} Barrat~A., Weigh~M., Eur. Phys. J. B, 2000,
        \textbf{13}, 547; 
        \bibdoi{10.1007/s100510050067}.

\bibitem{G} Georgii~H.-O.,  Gibbs measures and phase transitions.
de Gruyter Studies in Mathematics, {\bf 9}. Walter de~Gruyter \&
Co., Berlin, 1988.
\bibitem{KK} K\c{e}pa~D., Kozitsky~Y., Condens. Matter Phys., 2008, \textbf{11}, 313. 

\bibitem{JS} Jonasson~J, Steif~J.E., J. Theoret. Probab., 1999, \textbf{12},
549; 
\bibdoi{10.1023/A:1021690414168}.

\bibitem{Bovier} Bovier, A., Statistical mechanics of disordered systems.
A mathematical perspective. Cambridge Series in Statistical and
Probabilistic Mathematics, Cambridge University Press, Cambridge,
2006; \bibdoi{10.1017/CBO9780511616808.004}.

\bibitem{hincz} Hinczewski~M., Phys. Rev.~E, 2007,  \textbf{75},
06611104; 
\bibdoi{10.1103/PhysRevE.75.061104}.

\bibitem{modul} Kashtan~N., Mayo~A.E., Kalisky~T., Alon~U.,  PLoS Comput. Biol., 2009,
  \textbf{5}, e1000355; \\
  \bibdoi{10.1371/journal.pcbi.1000355}.


\end{thebibliography}
\end{document}